\documentclass{article}
\usepackage[margin=1.25in]{geometry} 
\usepackage{amsfonts}
\usepackage[cmex10]{amsmath}
\usepackage{mathtools, amsthm}
\usepackage{paralist}
\newcommand{\coloneqq}{=}
\newtheorem{theorem}{Theorem}
\newtheorem{lemma}[theorem]{Lemma}
\usepackage{hyperref}
\newcommand*{\mytitle}{Reconstruction and Estimation of Scattering Functions of Overspread Radar Targets}
\newcommand*{\myauthor}{Onur Oktay, G\"{o}tz Pfander, Pavel Zheltov}
\hypersetup{
pdfkeywords = {radar, channel estimation, spreading function, scattering function, time-varying operators},  
pdfauthor = {\myauthor},  
pdftitle = {\mytitle},
pdfpagelayout = OneColumn,
pdfstartview = FitH
}
\newcommand*{\epi}[1]{\,e^{2\pi i #1}}
\newcommand*{\empi}[1]{\,e^{-2\pi i #1}}

\newcommand*{\Z}{\mathbb{Z}}
\newcommand*{\C}{\mathbb{C}}
\DeclareMathOperator{\supp}{supp}

\newcommand*{\y}{y}   %echo
\newcommand*{\U}{U}   %Gabor submatrix
\newcommand*{\V}{V}   % inverse of \U D
\newcommand*{\x}{x}   %input signal
\renewcommand{\d}{\:\mathrm{d}}
\newcommand*{\EXP}{\mathbb{E}}
\newcommand*{\E}[1]{\EXP\left\{#1\right\}}
\newcommand*{\Srhat}{\widehat{S_r}}
\newcommand*{\Phat}[1]{\widehat{\Pi}_{{#1} J + r}}
\newcommand*{\conj}[1]{#1^{*}}
\newcommand*{\dnb}{{\textstyle\frac{n}{B}}}
\newcommand*{\dob}{\frac{1}{B}}

\DeclareMathOperator{\Var}{var}
\DeclareMathOperator{\Cov}{cov}
\DeclarePairedDelimiter\abs{\lvert}{\rvert}
\DeclarePairedDelimiter\absq{\lvert}{\rvert^2}

\DeclarePairedDelimiter\parensq{\lparen}{\rparen^2}
\DeclarePairedDelimiter\norm{\lVert}{\rVert}
\newcommand*{\itimes}{}
\newcommand*{\1}{\hspace{1.7em}}
\newcommand*{\2}{\1\1}

\begin{document}
\title{\mytitle}
\author{\myauthor
% \thanks{\today}%
\thanks{O.~Oktay, G.~E.~Pfander and P.~Zheltov are with Jacobs University Bremen}%
\thanks{Emails:\{o.oktay, g.pfander, p.zheltov\}@jacobs-university.de}%
\thanks{G.~E.~Pfander and P.~Zheltov acknowledge funding by the Germany Science Foundation (DFG) under Grant 50292 DFG PF-4, Sampling Operators.}
}
\maketitle
\begin{abstract}
In many radar scenarios, the radar target or the medium is assumed to possess randomly varying parts. The properties of a target are described by a random process known as the spreading function. Its second order statistics under the {WSSUS} assumption are given by the scattering function. 
Recent developments in the operator identification theory suggest a channel sounding procedure that allows to determine the spreading function given complete statistical knowledge of the operator echo. 
We show that in a continuous model it is indeed theoretically possible to identify a scattering function of an overspread target given full statistics of a received echo from a single sounding by a custom weighted delta train. Our results apply whenever the scattering function is supported on a set of area less than one. 
Absent such complete statistics, we construct and analyze an estimator that can be used as a replacement of the averaged periodogram estimator in case of poor geometry of the support set of the scattering function.
\end{abstract} 

\section{Introduction}\label{sec:intro}
In the classical delay-{D}oppler radar system, the echo $\y(t)$ that is reflected from a target can be expressed as a superposition of time-frequency shifts of the transmitted waveform $\x(t)$, that is,
\begin{equation}\label{eq:channel}
\y (t)= \iint \eta(\tau, \gamma)M_{\gamma}T_{\tau}\x(t) \d\tau \d\gamma,
\end{equation}
where $T_{\tau} \x(t) = \x(t-\tau)$ is a time shift operator, $M_{\gamma} \x(t) = \epi{\gamma t}\x(t)$ is a frequency shift operator and $\eta$
is the \emph{spreading function} of the target. If the target has randomly varying components, then $\eta(\tau, \gamma)$ is assumed
to be a random process. It is often assumed that the radar environment satisfies the wide-sense stationarity with uncorrelated scattering (WSSUS) assumption \cite{BelloChar,VanTrees}. In this case, $\eta(\tau, \gamma)$ satisfies the relation
\begin{equation}\label{eq:def.scat}
\E{ \conj{\eta(\tau, \gamma)} \eta(\tau',\gamma') } = \delta(\tau - \tau') \, \delta(\gamma - \gamma') \,C(\tau,\gamma),
\end{equation}
where $C(t,\gamma)$ is called the \emph{scattering function} of the target.

A problem in radar is to determine $C(t,\gamma)$ from the echo $\y$ \cite{Green,Harmon}. 
A standard approach is to view the scattering function $C(t,\gamma)$ as power spectral density of the 2-D stationary random process known as the \emph{time-variant transfer function}. This gives rise to the \emph{averaged periodogram estimator} of $C(t,\gamma)$.  One of the significant restrictions of this approach is the requirement to assume an \emph{underspread} target, that is, the support of $C(t,\gamma)$ must be contained in a rectangle of area less than one. 
Building up on the recent results in operator identification \cite{PfWal,KozPf,PRT08,Pfander1,GP08}, we show that it is the area of the support of $C(t,\gamma)$ itself that matters, and not the area of the bounding box. 
This fact calls for reclassification of the channels that would previously be considered overspread due to the non-rectangular geometry of the set $\supp C(t,\gamma)$. 

In Section \ref{sec:target_id} we prove the following theorem.
\begin{theorem}\label{thm:main}
Let $C(t,\gamma)$ be the scattering function of a radar target. Let the support of $C(t,\gamma)$ have area $M < 1.$ There exists $T>0$,  a natural number $J$ and a fixed $J$-periodic sequence $c_k,$ such that we can identify the scattering function $C(t,\gamma)$ from the received echo $\y(t) = H\x(t)$ where $\x(t) = \sum_{k\in\Z}c_k\delta_{kT}(t)$ is the sounding signal.
\end{theorem}

In the tradition of \cite{PfWalRecon}, we provide an explicit formula \eqref{eq:reconstruction} for the reconstruction of the scattering function that is supported on a possibly overspread domain of the time-frequency plane using a specially constructed delta train as the sounding signal. A procedure for the estimation of the scattering function is naturally derived from this reconstruction formula. 

The paper is organized as follows.  Section~\ref{sec:overview_radar} gives an overview of classical delay-{D}oppler radar. In Section~\ref{sec:target_id} we describe the target identification problem and prove the main result, \autoref{thm:main}. We suggest and analyze an estimator in Section~\ref{ssec:scat_func_est}, followed by our conclusions in Section~\ref{sec:conclusion}.

\section{Overview of radar}\label{sec:overview_radar}
The classical scenario in a delay-{D}oppler radar system is that a testing signal $\x(t)$  is transmitted. The signal
$\x(t)$ might be a short pulse, as well as a wideband linear chirp, coded waveform, pseudonoise sequence, etc. \cite{Sko80,Taylor2}.

The echo from a point target at distance $d$ and traveling at constant speed $v$, is considered to be of the form 
\[ \y = \eta_0 \x(t-t_0)\epi{t \gamma_0},\]
where $t_0 = 2d/c$, $c$ is the speed of light, $\gamma_0$ is the Doppler shift , and $\eta_0$ is the \emph{reflection coefficient} that depends on the distance and the speed of the point target. 

As is customary and indicated in \eqref{eq:channel}, we model the echo realized in a complete environment as a continuous superposition of time-frequency shifts of the transmitted signal~$\x(t)$ \cite{Green}. 
Equivalently to \eqref{eq:channel}, from the \emph{time-varying impulse response} 
\begin{equation}\label{eq:h}
h(t, \tau) \coloneqq \int \eta(\tau, \gamma)\epi{\gamma t} d\gamma
\end{equation}
we obtain the linear time-varying  representation of the channel
\begin{equation*}\label{eq:II.5}
\y(t) = \int h(t,\tau)\x(t-\tau) d\tau.
\end{equation*}

Certain characteristic features of a target or a radar scene, for example, sea clutter \cite{Gaarder,VanTrees,Doyuran,Fuhrmann}, are often modelled as random processes.
In this model, $h$ and $\eta$ are assumed to be a pair of random processes related via the one-sided Fourier transform \eqref{eq:h}.

Let the quantity
\[R_h(t,s;\xi,\eta) \coloneqq \E{\conj{h(t,\xi)}h(s,\eta)} \]
denote the \emph{autocorrelation function} of $h$, where $\E{\itimes}$ denotes  expected value. The usual WSSUS assumption in radar applications is that $h(t,\tau)$ is 
\begin{inparaenum}[\itshape a\upshape)]
\item zero-mean in both variables,
\item wide sense stationary (WSS) in $t$, and
\item has uncorrelated scatterers $\tau$,
\end{inparaenum}
which reads 
\begin{equation*} %\label{eq:WSSUS}
R_h(t,t+\tau;\xi,\eta) = P_h(\tau,\xi)\delta(\xi-\eta).
\end{equation*}		
Then the scattering function $C(t,\gamma)$ defined in \eqref{eq:def.scat} satisfies
\begin{equation*} %\label{eq:II.6}
 C(\tau,\gamma) \coloneqq \int P_h(t,\tau)\epi{t \gamma} \d t.
\end{equation*}
$C(t,\gamma)$ is the previously mentioned scattering function associated with the channel, and $P_h(t,\tau)$ is the channel's \emph{autocorrelation function}, or {ACF}. The computations above and in the remainder of this paper involve delta distributions, thus the convergence and equality of the integrals must be understood in a weak sense. We preserve current notation for clarity and relegate  a rigorous treatment of this matter to \cite{PfaZh}.

In radar applications, it is usually assumed that the scattering function $C(t,\gamma)$ of a target has its support on a time-frequency rectangle $[0, \Theta] \times [-\Omega/2, \Omega/2].$
The degree of dispersion of the echo is quantified by the \emph{spreading factor} $\Omega\Theta$. A target is said
to be \emph{underspread} if $\Omega\Theta < 1$ and \emph{overspread} if $\Omega\Theta > 1$. 
In recent years it has become increasingly clear that the correct criterion is that the area of the support of the scattering function must be less than one \cite{DBS10}. We show that it is indeed correct, and require only that $supp\, C(t,\gamma)$ is compactly supported within some rectangle $[0, \Theta] \times [-\Omega/2, \Omega/2]$ without demanding $\Omega\Theta <1$. 
A classical identification problem in radar and communications is to estimate the scattering function of a given
randomly varying target, or, equivalently, an linear time variant random channel. The general approach is to transmit a signal $\x(t)$, once or multiple
times, and to construct an estimator $\widehat{C}(t,\gamma)$ for the scattering function $C(t,\gamma)$ using the received echoes.
Obtaining measurements in this way is referred to as \emph{channel sounding} in communications. 

\section{Scattering function identification}\label{sec:target_id}
Let $\eta$ be the spreading function of a radar target such that $\supp C(t,\gamma) \subseteq [0,\Theta]\times[-\Omega/2, \Omega/2],$ with $\Omega\Theta$ possibly larger than 1. Let the set $\supp C(t,\gamma)$ be covered by the translations of the prototype rectangle with some $J$ prime number, as follows:
\[\supp C(t,\gamma) \subseteq \bigcup_{j=1}^J  R+(a_jT,b_jB)\]
where $a_j,b_j \in \Z$ index the cover, $T=\Theta/\Omega$, $B = \Omega/J\Theta$. This implies $JBT = 1$. Note that other bounding boxes, for example, a first quadrant box $[0,\Theta]\times[0,\Omega]$, can be accommodated by a simple translation; the case $JBT<1$ can be padded with empty boxes, if necessary. The theory of Jordan domain shows that for any compact set $M$ of area less than one the prime number $J$ and the covering $\{a_j, b_j\}_{j=1}^J$ can always be found. 
Given~$\eta$, we define patches of $\eta$ to be 
\begin{equation*}
 \eta_j(t,\gamma) \coloneqq \chi_R(t,\gamma) \, \eta(t+a_jT,\gamma+b_jB),
\end{equation*}
and their Fourier transforms 
\begin{equation*}
h_j(x,t) \coloneqq \int \eta_j(t,\gamma)\epi{\gamma x} \d\gamma.
\end{equation*}
We can now reconstruct
\[ \eta(t,\gamma) = \sum_{j=1}^J \eta_j(t-a_jT,\gamma-b_jB), \]
and
\begin{align*} 
  h(x,t) &= \sum_{j=1}^J \int \eta_j(t-a_jT,\gamma-b_jB)\epi{\gamma x} \d\gamma \\
   &= \sum_{j=1}^J \int \eta_j(t-a_jT,\gamma)\epi{(\gamma+b_jB) x} \d\gamma \\
   &= \sum_{j=1}^J \epi{x b_j B} \int \eta_j(t-a_jT,\gamma)\epi{\gamma x} \d\gamma \\
  &= \sum_{j=1}^J \epi{x b_j B} h_j(x,t-a_jT).
\end{align*}

For each $t$ fixed, $\eta_j(\itimes,t)$ is supported on $[0,B)$. Therefore, for any $\tau$ we have the orthonormal expansion  
\begin{equation}\label{eq:eta.orth.exp}\begin{split}
  \eta_j(t,\gamma) &= \dob\sum_{j\in\Z} \left( \int_0^B \eta_j(t,s)\epi{s (\tau + \dnb)} \d s\right) \empi{\gamma (\tau+\dnb)}
  \\
&= \dob\sum_{j\in\Z} h_j\left(\tau+\dnb,t\right)\empi{\gamma(\tau+\dnb)}.
\end{split}\end{equation}

In particular, setting $\tau  = \tau_j(t) = t+a_jT$, we obtain
\begin{align*}
  \eta(t,\gamma) &= \sum_{j=1}^J \eta_j(t-a_jT,\gamma-b_jB) \\
	&= \dob \sum_{n \in \Z}\sum_{j=1}^J h_j\left(t+\dnb,t-a_jT\right)\empi{(\gamma-b_jB)(t+\dnb)} \\
	&= \dob\sum_{n\in\Z}h\left(t+\dnb,t\right) \empi{\gamma (t+\dnb)}.
\end{align*}
For the random $\eta_j$'s, we have
\begin{align*} 
\EXP  &\: \{\conj{\eta_j(t,\gamma)}\eta_\ell (s,\lambda)\} \\
&= \delta\left(t-s+(a_j-t_\ell)T\right)\delta\left(\gamma-\lambda+(b_j-\gamma_\ell)B\right)  
	\chi_R(t,\gamma) \chi_R(s,\lambda) C(t+a_jT,\gamma+b_jB) \\
&= \delta\left(t-s+(a_j-t_\ell)T\right)\delta\left(\gamma-\lambda + (b_j-\gamma_\ell)B\right)
	\chi_R(t,\gamma)\chi_R(t+(a_j-t_\ell)T,\gamma+(b_j-\gamma_\ell)B) \\
&\1 \times C(t+a_jT,\gamma+b_jB).
\end{align*}
But this is zero unless $a_j=t_\ell$ and $b_j = \gamma_\ell$, that is $j=\ell$, in which case we have
\begin{equation*}
\E{ \conj{\eta_j(t,\gamma)}\eta_j (s,\lambda) } 
= \chi_R(t,\gamma)\delta(t-s)\delta(\gamma-\lambda)\itimes C\left(t+a_jT,\gamma+b_jB\right).
\end{equation*}
In particular, defining $C_j(t,\gamma) = \chi_R(t,\gamma)C(t+a_jT,\gamma+b_jB)$, we have the expansion 
\begin{equation}\label{eq:C.as.sum}
 C(t,\gamma) = \sum_{j=1}^J C_j(t-a_jT,\gamma-b_jB).
\end{equation}
Similarly, for the autocorrelation of $h$'s we have whenever $j \ne \ell$
\begin{equation*} 
\E{ \conj{h_j(\tau,t)}h_\ell (\upsilon,s) } = \iint \empi{(\gamma \tau -\lambda \upsilon)}\E{ \conj{\eta_j(t,\gamma)}\eta_\ell (s,\lambda) } \d\gamma \d\lambda = 0,
\end{equation*}
and, when $j=\ell$,
\begin{equation}\label{eq:III.10}\begin{split}
\E{\conj{h_j(\tau,t)}h_j (\upsilon,s)} &= \iint \empi{(\gamma \tau -\lambda \upsilon)} C_j(t,\gamma)\delta(t-s)\delta(\gamma-\lambda)\d\gamma \d\lambda \\
&= \int\empi{(\tau - \upsilon)} C_j(t,\gamma)\d\gamma\delta(t-s) \\
&= P_{h_j}(\tau-\upsilon,t)\delta(t-s),
\end{split}\end{equation}
where $P_{h_j}(\tau,t) \coloneqq\int \empi{\gamma \tau} C_j(t,\gamma)\d\gamma$.
Moreover, as a result of \eqref{eq:C.as.sum}, we have
\begin{equation}\label{eq:III.11}\begin{split}
P_h(\tau,t) &= \int \empi{\gamma \tau} C(t,\gamma)d\gamma \\
&= \sum_{j=1}^J \int \empi{\gamma \tau} C_j(t-a_jT,\gamma-b_jB)d\gamma \\
&= \sum_{j=1}^J \empi{\tau b_j B} P_{h_j}(\tau,t-a_jT),
\end{split}\end{equation}

Next, we derive a formula similar to \eqref{eq:eta.orth.exp} for the scattering functions $C_j(t,\gamma)$.
\begin{lemma}\label{lemma:III2}
With $\eta_j$, $h_j$, $P_{h_j}$ and $C_j$ defined as above, we have
\[C_j(t,\gamma) = \dob \sum_{n\in \Z} \epi{\gamma(\tau+n/B)}P_{h_j}\left(\tau+\dnb,t\right), \]
where $\tau$ is arbitrary.
\end{lemma}

\begin{proof}
 By equations \eqref{eq:eta.orth.exp} and \eqref{eq:III.10}, we have
\begin{align*}
\E{ \conj{\eta_j(t,\gamma)}\eta_j (s, \lambda)} 
&= \dfrac{1}{B^2}\sum_{n,m \in \Z} \epi{ (\gamma \tau +\gamma \dnb - \lambda \frac{m}{B})} \E{ \conj{h_j(\tau+\dnb,t)}h_j(\frac{m}{B},s) } \\
&= \dfrac{1}{B^2}\sum_{n,m \in \Z} \epi{(\gamma \tau +\gamma \dnb - \lambda \frac{m}{B})} P_{h_j}\left(\tau+\frac{n-m}{B},t\right)\delta(t-s) \\
&=\dfrac{1}{B^2}\sum_{n,m \in \Z} \epi{(\gamma \tau +\gamma \dnb - \lambda \frac{m}{B})} P_{h_j}\left(\tau+\dnb,t\right)\delta(t-s) \\
&=\dob\sum_{n \in \Z} e^{2 \pi i (\gamma \tau +\dnb)} P_{h_j}\left(\tau+\dnb,t\right)\delta(t-s)\delta(\gamma-\lambda).
\end{align*}
Since $\E{ \conj{\eta_j(t,\gamma)}\eta_j (s, \lambda) } = \delta(t-s)\delta(\gamma-\lambda)C_j(t,\gamma),$ the result follows.
\end{proof}

We are now ready to prove the main result of this paper.
\begin{proof}[Proof of \autoref{thm:main}]
Let $c_k$ be a $J$-periodic complex sequence. Transmit $\x(t) = \sum_{k\in\Z} c_k \delta_{kT}(t)$ and observe echo $\y(t) = H\x(t)$. Also, define
\begin{equation}\label{eq:III.12}
\y_n(t) \coloneqq (H\x)(t+nT)
= H\Big(\sum_k c_k \delta_{kT}\Big)(t+nT)
= \sum_{k\in\Z}c_k h(t+nT,t+(n-k)T),
\end{equation}
for $0\le t<T$ and $n \in \Z$. Clearly, if $h(t,\tau)$ is a stochastic process, so are~$\y_n(t)$. In fact, for $0 \leq t, \tau < T$ and $n \in \Z$, we have
\begin{align*} 
\E{ \conj{\y_n(t)}\y(\tau) } 
&= \sum_{k,l \in \Z} \conj{c_k}c_\ell \E{ \conj{ h(t+nT,t+(n-k)T)}h(\tau,\tau-\ell T) } \\
&= \sum_{k,l \in \Z} \conj{c_k} c_\ell P_h(nT+t-\tau, t+(n-k)T) \itimes\delta(t-\tau-(n-k+\ell )T).
\end{align*}
But for $0 \le t,\tau < T$,
\begin{equation*}
\delta\left(t-\tau-(n-k+\ell )T\right)= \left\{
	\begin{aligned}
		&\delta(t-\tau), \quad \text{if } n-k+\ell  = 0,  \\
		&0, \quad \text{if } n-k+\ell  \ne 0.
	\end{aligned}\right.
\end{equation*}
Thus
\begin{equation}\label{eq:Pi}\begin{split}
\E{ \conj{\y_n(t)}\y(\tau)} &= \sum_{k\in\Z} \conj{c_k} c_{k-n} P_h(nT,t+(n-k)T)\delta(t-\tau) \\
&= \sum_{k\in\Z} \conj{c_{k+n}} c_{k} P_h(nT,t-kT)\delta(t-\tau) \\
&= \Pi_n(t)\delta(t-\tau),
\end{split}\end{equation}
where we defined
\[\Pi_n(t) \coloneqq \sum_{k\in\Z} \conj{c_{k+n}} c_{k} P_h(nT,t-kT).\]
\newcommand*{\moB}{{\textstyle\frac{m}{B}}} 
Using \eqref{eq:III.11} and \eqref{eq:Pi}, we obtain
\begin{align*} 
\Pi_{r+mJ}(t) &= \sum_{k\in \Z} \conj{c_{k+r+mJ}}c_kP_h\left(\moB+rT,t-kT\right) \\
&= \sum_{k\in\Z}\conj{c_{k+r}}c_k \sum_{j=1}^J \empi{(\moB+rT)b_jB} 
P_{h_j}\left(\moB+rT,t-(k+a_j)T\right) \\
&= \sum_{j=1}^J \empi{r b_j / J} \itimes \sum_{k\in\Z} \conj{c_{k+r}}c_k
 P_{h_j}\left(\moB+rT,t-(k+a_j)T\right).
\end{align*}
Recall $JBT=1$ and denote for $r=1,\dotsc, J$ 
\[ 
S_r(t,\gamma) \coloneqq \dob \sum_{m\in\Z}\epi{\gamma T(mJ+r)}\Pi_{m J+r}(t).
\] 
By Lemma \ref{lemma:III2} we compute 
\begin{equation}\label{eq:def:S_r}\begin{split}
S_r(t,\gamma) &= \dob \sum_{m\in\Z}\epi{\gamma T(mJ+r)}\Pi_{m J+r}(t)   \\
   &= \sum_{j=1}^J \empi{r b_j/J} \sum_{k\in\Z} \conj{c_{k+r}}c_k \dob \sum_{m\in\Z} \epi{\gamma T(mJ+r)} 
 P_{h_j}\left((mJ+r)T,t-(k+a_j)T\right)    \\
  &= \sum_{j=1}^J \empi{r b_j/J} \sum_{k\in\Z}c^{*}_{k+r}c_k C_j(t-(k+a_j)T,\gamma).
\end{split}\end{equation}
But, $C_j(\tau,\gamma)\not\equiv 0$ only if $0 \le \tau <T$. Then, $C_j(t-kT-a_jT,\gamma)\neq 0$ if and only if
\[ t- T < (k+a_j)T \le t. \]
Since $0\le t < T$, we must have that $C_j(t-(k+a_j)T,\gamma) \ne 0$ only if $k+a_j = 0.$
Therefore, by \eqref{eq:def:S_r},
\begin{equation}\label{eq:III.16}
 S_r(t,\gamma) = \sum_{j=1}^J \empi{r b_j / J}c^{*}_{r-a_j}c^{\phantom{*}}_{-a_j}C_j(t,\gamma).
\end{equation}

Let $\{ c_{k,l} \}^{J-1}_{k,l=0}$ be the Weyl-Heisenberg frame for $\C^J$, where
\[c_{k,l}(r) = \empi{r l /J} c_{r-k},\ r = 1,\dotsc,J.\]
The coefficient sequence $\{c_k\}_{k=0}^J$ can be chosen so that any $J$ element subset of the so-called Gabor frame vectors is linearly independent \cite{KPR08}, a condition of the frame known as the \emph{Haar property}.
Moreover, the selection procedure allows us to choose $c$ to be unimodular. In fact, choosing the entries of $c$ randomly from a uniform distribution on a unit circle guarantees the Haar property will hold with probability one. In particular, this means that a matrix of our interest, determined by the geometry of the support set of $C(t,\gamma)$, is invertible. 
Let's denote by $\U$  the $J\times J$ matrix whose $j$-th column is $c_{a_j,b_j}$, that is, 
\[ \U = \left[ c_{a_1,b_1} \big\vert c_{a_2,b_2} \big\vert \dotsb \big\vert c_{a_j,b_j} \right].\]
Then, \eqref{eq:III.16} can be expressed in the matrix form as
\begin{equation*}\label{eq:III.17}
 [S_r(t,\gamma)]^{J}_{r=1} = \U^{*}[c_{-a_j}C_j(t,\gamma)]^{J}_{j=1}.
\end{equation*}
so that we can rewrite the above equation as 
\[ \left[S_r(t,\gamma)\right]_{r=1}^J = \U^{*}D\left[C_j(t,\gamma)\right]_{j=1}^J\]
with $D\in \C^{J\times J}$ a non-singular diagonal matrix with elements of $c$ on the diagonal, that is, $D_{ij} = c_{-a_j}\delta_{ij}$. Let $\V = [\V_{jr}]_{r,j=1}^J$ be the inverse of $\U^*D_c$. 
We can now recover $C_j(t,\gamma)$ pointwise from the autocorrelation of the output signal with an explicit reconstruction formula derived from \eqref{eq:III.12}, \eqref{eq:Pi} and \eqref{eq:def:S_r}, that is, for any $j \in [1,J]$ and $t\in[0,T), \gamma \in [0, B)$
\begin{equation}\label{eq:reconstruction}
C_j(t,\gamma) = \sum_{r=1}^J \V_{jr} S_r(t,\gamma) = 
\sum_{r=1}^J \V_{jr} \dob \sum_{m\in\Z}\epi{\gamma T(mJ+r)}\Pi_{m J+r}(t).
\end{equation}
where $\Pi_n(t)\delta(t-\tau) = \E{\conj{\y(t+nT)}\y(\tau)}.$
This completes the proof of \autoref{thm:main}.
\end{proof}

\section{Scattering function estimation}\label{ssec:scat_func_est}
Since $h(t,\tau)$ is a stochastic process, so is the returned echo $\y(t)$. 
Although \autoref{thm:main} enables us to reconstruct the scattering function perfectly from the stochastic process $\y(t)$, in practice we do not know the values of $\E{ \conj{\y_n(t)}\y(\tau)}$. Secondly, the delta trains that we use for identification of the channel have infinite energy, hence, so do their echoes. To address the first problem, we sound the channel repeatedly with the  same input signal $\x(t) = \sum_{k\in\Z} c_k \delta_{kT}(t)$ to obtain  $\y^{(1)}(t), \dotsc, \y^{(L)}(t)$, $L$ independent identically distributed samples of the channel output. As the channel is WSSUS, repeated soundings do not pose an insurmountable problem in general. Since the primary purpose of this paper is a proof of concept for the reconstruction of the scattering function with non-rectangular support, we do not explore the effects of time-gating the delta train. 

Regarding the second obstacle of infinite energy responses, we consider as data for our estimator the functions $\Pi_n(t)$ defined in \eqref{eq:Pi} for $t,\tau\in[0,T]$ as follows.  
\[ \Phat{m}(t)\delta(t-\tau) = \frac{1}{L} \sum_{l=1}^L \y^{(l)}(t-(mJ+r)T)^*\y^{(l)}(\tau). \]

Clearly, plugging the above into \eqref{eq:reconstruction} this induces an unbiased estimator $\widehat{C}(t,\gamma)$ of the scattering function $C(t,\gamma)$.  
\[ \widehat{C_j}(t,\gamma) = \sum_{r=1}^J \V_{jr} \dob \sum_{m\in \Z}\epi{\gamma T(mJ+r)}\Phat{m}(t). \]
We need the following trivial observation. We renumber the delta train 
\[ \x(t) = \sum_{k\in[J]} c_k^r \sum_{n\in\Z} \delta(t-(nJ-mJ + k-r)\]
and observe
\begin{align*}
\y & (t-(mJ+r)T) \\
 &= \iint \eta(\tau, \gamma)\epi{(t-(mJ+r)T)\gamma} \x(t-(mJ+r)T -\tau)\d\tau\d\gamma \displaybreak[3]\\
&= \sum_{k\in[J]} c_k^r \sum_{n\in \Z} \iint \eta(\tau, \gamma)\epi{(t-(mJ+r)T) \gamma} 
\delta(t -(mJ+r)T - (nJ-mJ-k-r)T- \tau )\d\tau\d\gamma \displaybreak[3]\\
&= \sum_{k\in[J]} c_k^r \sum_{n\in \Z} \int \eta(t -(nJ-k)T, \gamma) 
\epi{\gamma(t-((mJ+nJ-k+r)T)} \d\gamma \\
&= \sum_{k\in[J]} c_k^r \int \eta(t +k T, \gamma)\epi{\gamma(t-(mJ-k+r)T)} \d\gamma,
\end{align*}
where $n$ now depends on $k, m$ and $r$, and  $k$, as well as the sequence $c_k^r$, depend on $r$.

Let us pre-compute the covariance between $\Phat{m_1}(t)$ and $\Phat{m_2}(t)$. As different measurements $\y^{(l)}$ and $\y^{(l')}$  are independent, it is enough to consider
\begin{align*}
\EXP & \: \Phat{m_1}(t_1)\Phat{m_2}(t_2)^*\delta(t_1-t_3)\delta(t_2-t_4) \\
&= \frac{1}{L}\EXP \biggl\{ \y(t_1-(m_1J+r)T)^*\y(t_3)  \itimes \y(t_2-(m_2J+r)T)\y(t_4)^* \biggr\}  \displaybreak[3]\\
&= \frac1L \sum_{k_1, k_2,k_3,k_4 \in[J]} c_{k_1}^* c_{k_2}c_{k_3} c_{k_4}^* \sum_{n_1,n_2,n_3,n_4\in\Z} \iiiint \\
 &\1 \EXP \bigl\{ \eta\left(t_1-(n_1J-k_1)T, \nu_1\right)^* \eta\left(t_2-(n_2J-k_2)T, \nu_2\right) \\
 &\2 \times \eta\left(t_3-(n_3J-k_3)T,\nu_3\right)\eta\left(t_4 - (n_4J-k_4)T,\nu_4\right)^* \bigr\} \displaybreak[3] \\
 & \times \empi{[(t_1-(m_1J + n_1J-k_1 + r)T)\nu_1 + (t_4 - (n_4J-k_4)T)\nu_4]} \\
& \1\epi{[(t_3-(n_3J-k_3)T)\nu_3 + (t_2-(m_2J + n_2J-k_2+r)T)\nu_2]} \d\nu_1 \d \nu_2 \d\nu_3 \d \nu_4  
\end{align*}
To estimate the variance, we assume that all $\eta(t,\gamma)$ are jointly proper Gaussian processes. With the help of the Isserlis' moment theorem, we simplify the expectation 
\begin{align*}
\EXP &\bigl\{ \eta\left(t_1-(n_1J-k_1)T, \nu_1\right)^* \eta\left(t_2-(n_2J-k_2)T, \nu_2\right) \\ 
&\2 \eta\left(t_3-(n_3J-k_3)T,\nu_3\right)\eta\left(t_4 - (n_4J-k_4)T,\nu_4\right)^* \bigr\} \displaybreak[3]\\
&= C\bigl(t_1 - (n_1 J-k_1)T, \nu_1\bigr)\delta\bigl(t_1 - (n_1 J-k_1)T-(t_2 - (n_2J-k_2)T)\bigr)\delta(\nu_1-\nu_2) \\
&\2\times C\bigl(t_3-(n_3J-k_3)T,\nu_3\bigr)\delta\bigl(t_3-(n_3J-k_3)T-(t_4 - (n_4J-k_4)T)\bigr) \delta(\nu_3-\nu_4) \\
&\1 +  C\bigl(t_1 - (n_1 J-k_1)T, \nu\bigr) \delta\bigl(t_1 - (n_1 J-k_1)T - (t_3-(n_3J-k_3)T)\bigr) \delta(\nu_1 - \nu_3) \\
&\2\times C\bigl(t_2 - (n_2J-k_2)T, \nu_2\bigr)\delta\bigl(t_2 - (n_2J-k_2)T - (t_4 - (n_4J-k_4)T)\bigr) \delta(\nu_2 - \nu_4) \\
&\1 + C\bigl(t_1 - (n_1J-k_1)T, \nu_1\bigr)\delta\bigl(t_1 - (n_1J-k_1)T -(t_4 - (n_4J-k_4)T)\bigr) \delta(\nu_1 - \nu_4) \\
&\2\times C\bigl(t_2 - (n_2J-k_2)T,\nu_2\bigr)\delta\bigl(t_2 - (n_2J-k_2)T- (t_3-(n_3J-k_3)T)\bigr)\delta(\nu_2 -\nu_3). 
\end{align*}
All three summands can be estimated similarly.  We provide detail for the first summand only. Considerations of delta functions and support constraints of the scattering function cause cancellations. In particular, we have $n_1 = n_2 = 0$, $k_1=k_2$, $n_3=n_4 =0$, $k_3=k_4$, 
\begin{align*}
I_1(t) =& \sum_{k_1, k_2,k_3,k_4 \in[J]} c_{k_1}^* c_{k_2}c_{k_3} c_{k_4}^* \sum_{n_1,n_2,n_3,n_4\in\Z} \iiiint  \d\nu_1 \d \nu_2 \d\nu_3 \d \nu_4 \\
& \times C\bigl(t_1 - (n_1 J-k_1)T, \nu_1\bigr)\delta\bigl(t_1 - (n_1 J-k_1)T-(t_2 - (n_2J-k_2)T)\bigr) \delta(\nu_1-\nu_2) \itimes \\
& \times C\bigl(t_3-(n_3J-k_3)T,\nu_3\bigr)\delta\bigl(t_3-(n_3J-k_3)T-(t_4 - (n_4J-k_4)T)\bigr) \delta(\nu_3-\nu_4)  \\
& \times \epi{[-(t_1-(m_1J + n_1J-k_1 + r)T)\nu_1 + (t_2-(m_2J + n_2J-k_2+r)T)\nu_2 + (t_3-(n_3J-k_3)T)\nu_3 - (t_4 - (n_4J-k_4)T)\nu_4]} \\
=& \sum_{k_1, k_3 \in[J]} c_{k_1}^* c_{k_1}c_{k_3} c_{k_3}^* \sum_{n_1, n_3 \in\Z} 
\iint C\bigl(t_1 +k_1T, \nu_1\bigr) \empi{(t_1-t_2)\nu_1} C\bigl(t_3+k_3T,\nu_3\bigr) \\
& \times  \empi{(m_1- m_2) J  T \nu_1}  \d\nu_1 \d\nu_3. \\
\end{align*}
Hence, we have for the covariance 
\[\Cov\left(\Phat{m_1}(t),\Phat{m_2}(t)\right) = I_1(t) + I_2(t) + I_3(t).\]
We can now estimate the variance on $(t,\gamma) \in [0,T)\itimes [0,B)$. 
\begin{align*}
\bigl\vert\Var & \: \Srhat(t,\gamma)\bigr\vert  
= \abs*{\Var \dob \sum_{m} \epi{\gamma T(mJ+r)}\Phat{m}(t) } \displaybreak[3]\\
&= \frac{1}{B^2} \abs*{\sum_{m_1,m_2} \epi{\gamma T J (m_1-m_2)} \Cov \Phat{m_1}(t)\Phat{m_2}(t)^*} \displaybreak[3]\\
&= \frac{1}{B^2} \abs*{\sum_{m_1,m_2}\epi{\gamma T J (m_1-m_2)} \left(I_1(t) + I_2(t) + I_3(t)\right)} \displaybreak[3]\\
&= \frac{1}{B^2} \Big\lvert
\sum_{{k_1\in[J]}} \absq{c_{k_1}} \int C\bigl(t+k_1T, \nu_1\bigr)  \sum_{m_1} \epi{(\gamma -\nu_1)T J m_1}   \d\nu_1 \\
&\1\times  \sum_{k_3\in[J]} \absq{c_{k_3}} \int  C\bigl(t+k_3T,\nu_3\bigr)  \sum_{m_2} \epi{(\nu_3 - \gamma)T J m_2} \d\nu_3  \displaybreak[3] \\
&+\sum_{{k_1\in[J]}} \absq*{c_{k_1}} \int C\bigl(t+k_1T, \nu_1\bigr) \sum_{m_1}  \epi{[(\gamma+\nu_1) m_1 J T + r T\nu_1]}\d\nu_1 \\
&\1 \times \sum_{{k_2\in[J]}} \absq*{c_{k_2}} \int  C\bigl(t+k_2 T, \nu_2\bigr) \sum_{m_2} \empi{[(\gamma+\nu_2) m_2 J T + r T \nu_2]}\d \nu_2  \displaybreak[3] \\
&+ \sum_{{k_1\in[J]}} (c_{k_1}^*)^2  \int C\bigl(t+k_1 T, \nu_1\bigr) \sum_{m_1} \epi{(\gamma -2\nu_1)T J m_1} e^{-4\pi i \nu_1 (t+k_1 T- r T)} \d \nu_1 \\
&\1 \times  \sum_{{k_2\in[J]}} c_{k_2}^2  \int C\bigl(t+k_2 T, \nu_2\bigr)  \sum_{m_2} \empi{(\gamma-\nu_2) T J m_2}e^{4\pi i \nu_2 (t- k_2 T -r T)} \d \nu_2 \Big\rvert \displaybreak[3]\\
&=\frac{1}{B^2}\biggl\vert 
\bigl(\sum_{{k\in[J]}} \absq{c_k} C(t+k T,\gamma)\Bigr)^2  + \Bigl(\sum_{{k\in[J]}} \absq{c_k} C(t+k T,B-\gamma)\Bigr)^2 \\
&\1 + \Big\lvert \sum_{{k\in[J]}} \parensq{c_k} \bigl(C(t+kT,\gamma/2)  + C(t+kT, (B+\gamma)/2) \bigr) \Big\rvert^2 \biggr\vert \\
&\leq \frac{J}{B^2}\sum_{{k\in[J]}} \absq*{C(t+kT,\gamma)} +\absq*{C(t+kT,B-\gamma)} \\
&\1 + 2\absq*{C(t+kT,\gamma/2)} + 2\absq*{C(t+kT(B+\gamma)/2)}.
\end{align*}

This gives us a pointwise estimate. We now average to obtain the following estimate for the variance of $\widehat{C_j}$,
\begin{equation*}
\abs*{\Var \widehat{C_j}}  = \frac{1}{BT}\iint \abs*{\V \Var \Srhat(t,\gamma)  \V^*} \d t\d\gamma \leq  \frac{4\norm{\V}^2 J^2}{LB^2} \norm{C}_2^2.
\end{equation*}
This completes the bias-variance analysis. 
\section{Conclusion}\label{sec:conclusion}
In this paper, we have shown that it is possible to recover the scattering function of a target 
using the second order statistics of the returned echoes from sounding by a custom weighted delta train, provided that $\supp C(t,\gamma)$ is contained in a union of rectangles of total area 1. This includes channels commonly considered overspread under the classic criterion that the area of the bounding box is less than 1. 
In the abstract setting of a continuous {WSSUS} channel sounded by an infinite delta train, being given complete second-order statistics of the received signal, we guarantee an exact recovery of the scattering function. 
It is interesting to note that a single signal suffices for recovery of an entire collection of channels which share the gridding of the support set of their scattering functions, but not the support itself.  
We suggest that the proposed technique to reshuffle the patches of the support of the scattering 
function can and should be seen as a replacement of the Zak transform in the efficient 
implementation of the averaged periodogram estimator procedure offered  in \cite{Matz2}. 

For completeness, we give an explicit recipe of a simple unbiased estimator, with a predictable behavior of the variance given repeated soundings under the assumption of pointwise jointly proper Gaussianity of the spreading function. 
It would be interesting to explore the stability of the technique under the time-gating of the input signal. 
We currently do not pursue this direction, since the purpose of this paper is to establish the possibility of the scattering function reconstruction and to demonstrate the usefulness of the weighted delta train technique in the field of radar and sonar acquisition. 
\bibliographystyle{alpha}
 \bibliography{ScatteringFunction}
\end{document}